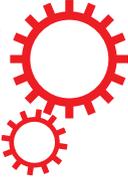

# Capturing relativistic wakefield structures in plasmas using ultrashort high-energy electrons as a probe

C. J. Zhang[1,2,4], J. F. Hua[1], X. L. Xu[3], F. Li[1], C.-H. Pai[1], Y. Wan[1], Y. P. Wu[1], Y. Q. Gu[2], W. B. Mori[3], C. Joshi[3] & W. Lu[1,4]

A new method capable of capturing coherent electric field structures propagating at nearly the speed of light in plasma with a time resolution as small as a few femtoseconds is proposed. This method uses a few femtoseconds long relativistic electron bunch to probe the wake produced in a plasma by an intense laser pulse or an ultra-short relativistic charged particle beam. As the probe bunch traverses the wake, its momentum is modulated by the electric field of the wake, leading to a density variation of the probe after free-space propagation. This variation of probe density produces a snapshot of the wake that can directly give many useful information of the wake structure and its evolution. Furthermore, this snapshot allows detailed mapping of the longitudinal and transverse components of the wakefield. We develop a theoretical model for field reconstruction and verify it using 3-dimensional particle-in-cell (PIC) simulations. This model can accurately reconstruct the wakefield structure in the linear regime, and it can also qualitatively map the major features of nonlinear wakes. The capturing of the injection in a nonlinear wake is demonstrated through 3D PIC simulations as an example of the application of this new method.

Plasma based accelerators have recently generated several gigaelectronvolts, high quality electron and positron beams in a short distance with a high efficiency[1–6], making them promising candidates for compact light sources[7–9] and future colliders[10]. Such accelerators can be driven either by intense laser pulses[1–4] or relativistic charged particle beams[5,6,11,12]. While a great deal of key physics of plasma-based wakefield accelerators[13] has been verified experimentally, other processes such as evolution of the wake (excitation, propagation and damping) and injection dynamics are still under investigation. Compared with theory[13,14] and the widely-used particle-in-cell (PIC) simulations[15,16], direct experimental observation of the field structure is scant due to the difficulty of directly probing the wake.

Direct measurements of the plasma wakefield are extremely challenging because these wakes are highly transient (~ps lifetime), microscopic (wavelengths and widths from a few μm to few hundred μm) and propagate at nearly the speed of light. Many efforts have been made to meet this challenge using optical probes. Frequency-domain-interferometry (FDI) was used to reconstruct density structures of the wake first in 1D and then in 2D[17–19]. Shadowgraphs of plasma wakes were taken by using an ultrashort laser probe[20], and the quality of the shadowgraph was recently improved by using a probe with duration less than 6 fs[21]. However, these optical methods invariably rely on the change of refractive index that is proportional to the plasma density variations. Therefore, these optical probing methods have not been demonstrated yet for diagnosing wakes in low density plasmas (~$10^{17}$ cm$^{-3}$) suitable for building a 10 GeV plasma wakefield accelerator stage[14].

It is desirable to directly probe the electric field structure instead of the density variations because the field is responsible for the acceleration and focusing of the injected bunch. To date, researchers have utilized charged particle beams including protons and electrons to visualize macroscopic electric and/or magnetic field structures in plasmas[22–26] but not relativistic wakes moving at nearly the speed of light. In order to spatially and temporally resolve relativistic wakefields, a charged particle probe beam capable of giving a spatial resolution as small as a few μm and a temporal resolution of as small as a few fs is necessary. These rigorous constraints can be met if a few fs long, relativistic electron bunch is used as a probe. Such a probe can in principle be generated from a laser wakefield accelerator (LWFA)[27].

[1]Department of Engineering Physics, Tsinghua University, Beijing 10084, China. [2]Laser Fusion Research Center, China Academy of Engineering Physics, Mianyang, Sichuan 621900, China. [3]University of California, Los Angeles, California 90095, USA. [4]IFSA Collaborative Center, Shanghai Jiao Tong University, Shanghai 200240, China. Correspondence and requests for materials should be addressed to W.L. (email: weilu@tsinghua.edu.cn)

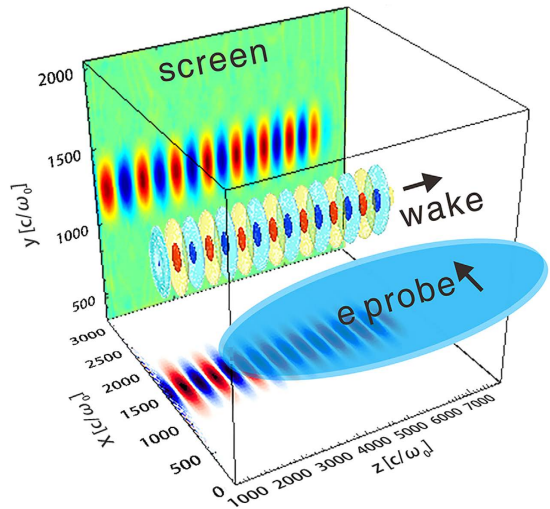

**Figure 1. Conceptual illustration.** A plasma wake propagating along the *z* direction can be probed by an electron bunch propagating along the *x* direction. After traversing through the wakefield, the probe will form a density pattern on a screen placed in the *yz* plane at an appropriate distance, due to the momentum modulation caused by the wakefield. The wakefield structure can consequently be reconstructed by analyzing the density modulation. The unit $c/\omega_0$ equals to 0.127 μm, where $c$ is the light speed in vacuum and $\omega_0$ is the frequency of the 800 nm laser.

In this paper, we propose a new method of using a femtosecond relativistic electron probe (FREP) to directly map out the plasma based wakefield structure. The validity and usefulness of this method for capturing the wakefield structure and its evolution is demonstrated through 3D PIC simulations. A theoretical model is developed to connect the transverse momentum modulation of the probe introduced by the wakefield to the density modulation of the probe. By analyzing the measured density modulation, the structure of a linear wakefield is accurately reconstructed. The qualitative reconstruction of a nonlinear wake is also presented. The effects of probe characteristics, including probe pulse length, energy spread, emittance, divergence and charge are discussed, showing that this method is feasible by using realistic probes. As an example of potential applications of this new tool, imaging of the downramp injection in a nonlinear wake using FREP is presented.

## Results

**Snapshot of Wakefield.** The concept behind probing the relativistic wake using a high-energy electron bunch is as follows: As the electron probe traverses through the wakefield, its transverse (relative to the propagation direction) momentum will be modulated by the electric field of the wake. This momentum modulation will evolve into a density modulation after a short drift distance. If a screen is placed at an appropriate diatance, a clear snapshot of the wakefield can be recorded (as shown in Fig. 1). The position of the screen is variable in experiment (a few to hundreds of mm) depending on the wakefield strength and the energy of the probe beam.

We first show the concept through PIC simulations using the code OSIRIS[28] where the wakes are probed by an ideal electron beam, i.e., with zero energy spread (central energy ~200 MeV), zero emittance, flat-top current profile and very short pulse length (6.8 fs, limited by the grid size of the simulation box). Figure 2 shows 3D PIC simulation snapshots of two different wakes, namely a linear wake (a) and a 3D nonlinear wake (c). The details of the simulation are described in the Methods section. Figure 2(b,d) give the probe density modulation recorded by a virtual screen placed behind the plasma, at a distance of 20 mm and 1 mm respectively. It can be clearly seen from these simulations that many basic informations regarding the wake structures, such as the wake period, width and shape, can be obtained from these snapshots directly. By varing the time delay between the probe and the driver, the evolution processes of the wakefield can also be obtained. Therefore this method could be a powerful tool to the study of wakfield physics, especially for the cases relevant to high energy physics applications, where very high power laser or electron/positron beams are used to drive a very low density plasma wake.

Beside the qualitative inforamtion that one can directly obtain from the above density snapshots, it will be very useful if the fields can be reconstructed from these density variations. In the next section, we will discuss how quantitative information of the fields can be obtained from the electron probe snapshots such as one shown in Fig. 2(b), i.e., for the case of linear wakes.

**Model for reconstructing the wakefield.** To quantitatively reconstruct the field structure, a theoretical model is developed. The model assumes that the wake is quasi-static, which means that the evolving time of the wake is much longer than the probe transition time. The probe transition time is proportional to the width of the wake, which is close to the plasma wavelength for typical wakes. Usually the evolving length of the wake is much longer than the plasma wavelength, especially in the case of linear wakes. Therefore the quasi-static asumption can be easily satisified. We also assume that the wakefield is cylindrically symmetric and propagates along the $z$ direction, and the probe traverses orthogonally along the $x$ direction, as shown in Fig. 1. The deflection angle of an electron can thus be expressed as:

$$\vec{\theta}(y, z) = \int_{-s}^{s} \frac{-e\vec{E}(x, y, z - \beta_E x)}{\beta c p_0} dx \quad (1)$$

where $\beta$ is the speed of the probe in unit of $c$, $p_0$ is the initial momentum of the probe, $\beta_E$ is the phase velocity of the wakefield and $\vec{E}(x, y, z)$ is the electric field of the wake being probed that contains both the longitudinal $E_z$ and the transverse $E_r$ components. The integral limits $\pm s$ should be large enough to cover the transverse extent of the wakefield. Here $p_0 \gg m_e c$ and the momentum modulation $\delta \vec{p}_{x,y,z} \ll p_0$ are assumed so that we neglect the mometum modulation along the direction of propagation of the probe. In Eq. (1), the magnetic field is neglected for simplicity. This assumption is valid for linear wakes where $B_\theta \ll E_{z,r}$. For nonlinear wakes, the magnetic field may be comparable with the electric field, however, the integration of the electric field, $\int E_z dx$ is still much larger than that of the magnetic field, $\int B_\theta dx$. As a consequence, Eq. (1) is still a good approximation in most cases. The detailed comparison of the contribution from the electric field and the magnetic field is presented in the supplemetary materials.

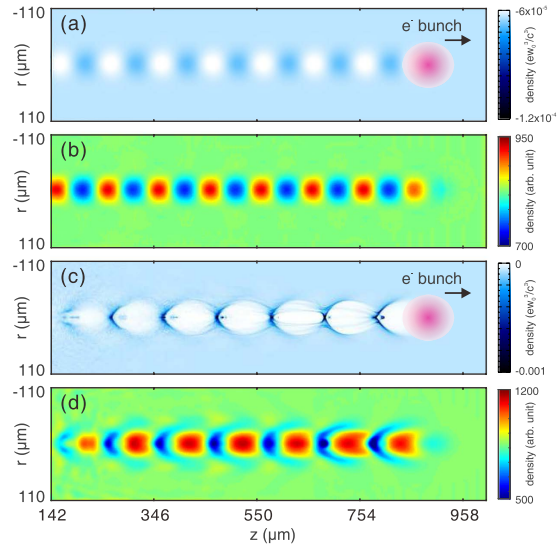

**Figure 2. Snapshots of wakefield.** (**a**) shows the plasma density map of a linear wake and (**b**) shows the corresponding electron probe snapshot. (**c,d**) show the plasma density map and electron probe snapshot for a nonlinear wake respectively. The electron drive bunches that drive the wakes are also shown in (**a,c**).

This deflecting angle leads to a displacement of $\vec{d} = \vec{\theta} L$ after the probe electrons propagate a distance $L$ in free space after exiting the wake. The self-wake and the space charge effects of the probe beam are negligble under the assumptions that the charge density of the probe is much lower than the plasma density and the energy of the probe electrons is suficiently high. If the displacement is small enough, trajectory crossing between the probe electrons will not occur, therefore a fluid element analysis is valid. Using fluid analysis, the density perturbation (defined as $\delta n/n_0 = (n - n_0)/n_0$, here $n_0$ is the initial density) can be written as:

$$\frac{\delta n}{n_0} = -\nabla \cdot \vec{d} \quad (2)$$

Combining Eq. (1) with (2), the density perturbation can be related to the field directly,

$$\frac{\delta n}{n_0}(y, z) = \nabla \cdot \frac{eL}{\beta c p_0} \int_{-s}^{s} \hat{z} E_z(r, z - \beta_E x) + \hat{y} \frac{y}{r} E_r(r, z - \beta_E x) dx \quad (3)$$

It can be seen from Eq. (3) that the density perturbation at a specific point $(y, z)$ is due to the combined contribution from $E_z$ and $E_r$. This can be decoupled by using Panofsky-Wenzel theorem[29] for accelerating cavities. The Panofsky-Wenzel theorem relates the transverse variation of the longitudinal field to the longitudinal variation of the transverse field by $\partial E_z/\partial r = \partial (E_r - B_\theta)/\partial (z - ct)$. In the linear regime, the magnetic field can be neglected since $B_\theta \ll E_r$, this theorem thus can be rewritten as $\partial E_z/\partial r = \partial E_r/\partial (z - ct)$. Taking the gradient of Eq. (3) along $z$ direction, we have

$$\frac{\partial I}{\partial z} = \kappa \nabla^2 \int_{-s}^{s} E_z(r, z - \beta_E x) dx \quad (4)$$

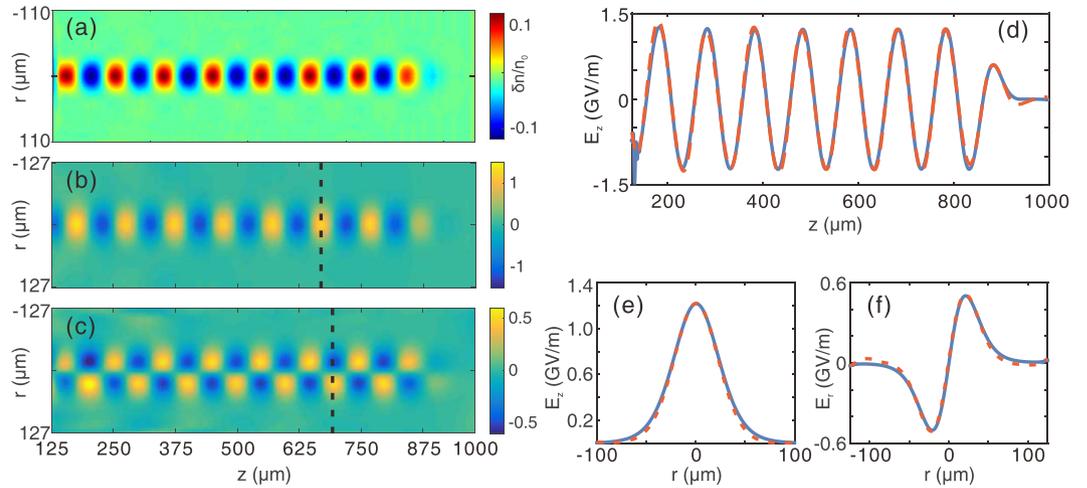

**Figure 3. Reconstruction of a linear wakefield.** (**a**) Probe density perturbation as recorded on a virtual screen. (**b,c**) are the reconstructed $E_z$ and $E_r$ field components, respectively. (**d**) On-axis ($r = 0$) lineout of reconstructed (red dashed line) and simulated (blue solid line) $E_z$. (**e,f**) show the transverse lineout of $E_z$ and $E_r$ at the positions indicated by the black dashed line in (**b,c**), respectively. The blue solid lines are for the simulated fields, while the red dashed lines are for the reconstructed fields.

where $\kappa = eL/\beta c p_0$, $I = \delta n/n_0$ and $\partial E_z/\partial r = \partial E_r/\partial z$ is applied. Similarly, the relation between density perturbation and the $E_r$ field can be written as

$$\frac{\partial I}{\partial y} = \kappa \nabla^2 \int_{-s}^{s} \frac{y}{r} E_r(r, z - \beta_E x) dx \tag{5}$$

Although the density perturbation contributions from $E_z$ and $E_r$ are now decoupled, Eqs (4) and (5) can not be solved directly, since the integrand changes with $x$ (time), which is a natural result of the motion of the wake. Thus further simplification is required. In the linear regime, the wakefield has the form of $E_z \sim \exp(-r^2/2\sigma_E^2) \cos k_p z$ and $E_r \sim r \exp(-r^2/2\sigma_E^2) \sin k_p z$ for a driver with transverse Gaussian profile, where $k_p$ is the plasma wave number and $\sigma_E$ is the width of the wake. Given that the wake is relativistic ($\beta_E \sim 1$), the integrands in Eqs (4) and (5) can be expressed by the integral of a static field with the same form multiplied by an average factor of $K_m = \exp(-(k_p \sigma_E)^2/2)$, as can be verified by Eq. (6):

$$\int_{-\infty}^{\infty} e^{-x^2/2\sigma_E^2} \cos k_p(z - x) dx = e^{-(k_p \sigma_E)^2/2} \int_{-\infty}^{\infty} e^{-x^2/2\sigma_E^2} \cos k_p z dx \tag{6}$$

This equation shows that the blurring effect caused by the propagation of the wakefield for an infinitely short probe can be removed for linear wakes with specific field distributions.

Under the above assumptions, the equations for reconstructing the wakefield can be summarized as

$$\frac{\partial I}{\partial z} = \kappa K_m \nabla^2 \int_{-s}^{s} E_z'(r, z) dx \tag{7}$$

and

$$\frac{\partial I}{\partial y} = \kappa K_m \nabla^2 \int_{-s}^{s} \frac{y}{r} E_r'(r, z) dx \tag{8}$$

where $\kappa = eL/\beta c p_0$ and $I = \delta n/n_0$ as before and $K_m = e^{-(k_p \sigma_E)^2/2}$. $E_z'$ and $E_r'$ are the corresponding static field with the same form as $E_z$ and $E_r$. The averaging effect caused by the motion of the wake, which depends on the width and period of the wake, is represented by the factor $K_m$. In an experiment, the energy of the probe $p_0$ and the drift distance $L$ are known. Furthermore, the width of the wakefield $\sigma_E$ and the probe density modulation $n$ can be measured from the snapshot in a single shot. The unperturbed probe density $n_0$ can be measured separately by taking a reference shot or be constructed from the data of $n$ by assuming that the initial probe density is continuous and smooth. With these information in hand, the field can be reconstructed by solving Eqs (7) and (8) with the following three steps: First, taking the partial derivative of the measured dnesity perturbation $I$ to get $\partial I/\partial z$ and $\partial I/\partial y$. Then sovling the Poisson equation to get the integral terms in Eqs (7) and (8). These integral terms are Abel transforms of the integrands $E_z'(r, z)$ and $E_r'(r, z)/r$, thus can be solved by performing the standard Abel inversion[30].

**Reconstruction of a linear wakefield.** Figure 3 shows the reconstruction results of a linear wake. The details of the simulation are described in the Methods section. Figure 3(a) shows the probe density perturbation on the virtual screen, which is placed 20 mm behind the plasma. Since the wake being probed is linear, the modulated density has a sinusoidal distribution along the longitudinal direction $z$ and approximately a Gaussian distribution along the transverse direction $y$. Figure 3(b,c) show the reconstructed $E_z$ and $E_r$ field components respectively. Peaks of density perturbation on the screen correspond to positions where the gradient of electric field has a maximum or minimum as expected from Poisson's equation. It can also be seen that $E_z$ is out of phase by $\pi/2$ radians compared with $E_r$ everywhere along the $z$ axis, a typical feature of linear wakes. The axial lineout of $E_z$ is shown in Fig. 3(d), where the red dashed line represents the reconstructed field, which agrees very well with the simulated field (the blue solid line). Figure 3(e,f) show the radial lineout of $E_z$ and $E_r$ at the positions indicated by the black dashed line in (b) and (c) respectively. The radial variation of $E_r$ shows that $E_r$ is zero on axis and changes sign as one goes through the axis as expected.

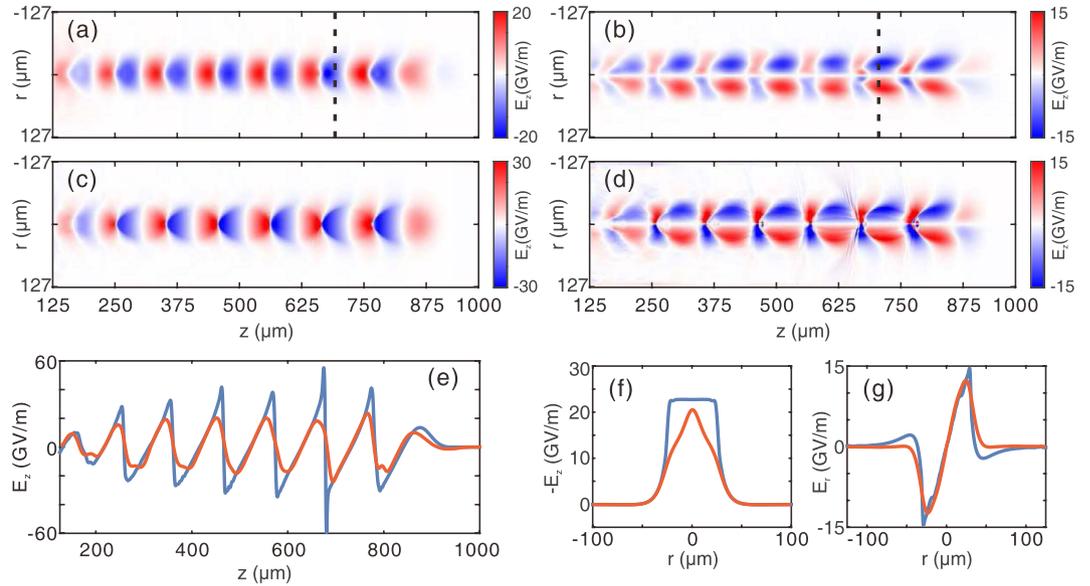

**Figure 4. Reconstruction of a nonlinear wakefield.** (**a,b**) are the reconstructed $E_z$ and $E_r$ fields respectively for a nonlinear wake. The corresponding probe density map is shown in Fig. 2(d). The simulated $E_z$ and $E_r$ fields are shown in (**c,d**), respectively. Axial lineouts of $E_z$ are shown in (**e**), where the red solid line represents the reconstructed field and the blue solid line is for the simulated field. In (**f,g**), the transverse lineout of $E_z$ and $E_r$ are shown, with the same color definition as in (**e**).

Figure 3 is a case where the $E_z$ field has a sinusoidal form and the $E_r$ field has a Gaussian form, for which the theoretical model is rigorously valid. But the usefulness of the method is not limited to this kind of wakes. Actually, this method can give very good reconstruction as long as the wake does not deviate too much from the wake shown in Fig. 3. As an example, the reconstruction results of a linear wake generated by an electron driver with super-Gaussian transverse profile are presented in the supplementary materials.

**Reconstruction of a nonlinear wakefield.** As the wakefield becomes strongly nonlinear, the longitudinal field $E_z$ starts to resemble a saw tooth function. For the 3D nonlinear blowout regime, a near hollow plasma cavity is formed by the strong transverse field of the electron beam driver or by the ponderomotive force of an intense laser driver, leading to a constant $E_z$ field on the transverse dimensions and a focusing $E_r$ field varying linearly with $r$[14]. Furthermore, the magnetic field becomes comparable with the $E_r$ field in a nonlinear wake. As a consequence, the application of Eqs (7) and (8) can only give qualitative reconstruction of a nonlinear wake. For nonlinear wakes, the averaging factor $K_m = e^{-(k_p \sigma_E)^2/2}$ in the linear case need to be modified. Eq. (7) now becomes $\frac{\partial I}{\partial z} = \kappa \nabla^2 \int_{-s}^{s} E_z(r, z - \beta_E x)dx \approx \kappa K_m \nabla^2 \int_{-s}^{s} E_z'(r, z)dx$ with a new averaging factor $K_m = 2e^{-((k_p \sigma_E)^2 - 1)/2}$ (for $k_p \sigma_E \geq 1$), which is an approximation based on simulations. We note that the corresponding static field $E_z'$ only maps the relatively low-frequency components of $E_z$, with the high frequency parts almost fully averaged out. Since the transverse distribution of $E_z$ of a nonlinear wake can no longer be approximated by a Gaussian function, here $\sigma_E = (2/\pi)^{1/2} r_b$ is used to calculate the equivalent width of the wakefield, where $r_b$ is the radius of the bubble[14].

The qualitative reconstruction of a nonlinear wakefield corresponding to the density map in Fig. 2(d) is shown in Fig. 4. Figure 4(a,b) show the $(r, z)$ plots of the reconstructed longitudinal and transverse fields respectively. The simulated longitudinal and transverse fields are also shown in Fig. 4(c,d). First, one can see that there is a close similarity on the wake structure and amplitude between the reconstructed and the actual field distributions, indicating that such a method can still be very useful for nonlinear wakes. However, the on-axis $E_z$ fields in Fig. 4(e) clearly show that the reconstructed field is much smoother than the real field, and this is due to the natural suppression of the high-frequency components. To see how this happens, we can approximate the $E_z$ field as a saw tooth function that is a sum of sinusoidal functions with harmonic of the plasma wave number $k_p$. The density perturbation of the probe is indeed the summation of all the harmonic components. However, for higher order harmonics with wave number $nk_p$, the averaging factor will be $e^{-(nk_p\sigma_E)^2/2}$ in Eq. (6), which becomes much smaller than the fundamental mode factor $K_m = e^{-(k_p\sigma_E)^2/2}$. Therefore, the information contained in the high-frequency components is significantly suppressed, and the reconstructed field becomes smoother than the actual waveform of the wake. Similar explanations can be adopted for understanding the differences between the transverse distributions of the reconstructed (red line) and the original $E_z$ and $E_r$ fields (blue line) shown in Fig. 4(f,g). The corresponding longitudinal positions of these lineouts are indicated by the dashed black lines in Fig. 4(a,b). Here we emphasize that even though the high-frequency components are lost in the reconstruction, the main features of the wake, such as the wake period, wake width and the overall shape remain close to the original ones, justifying the usefulness of this method as an important tool for the understanding of nonlinear wakefields.

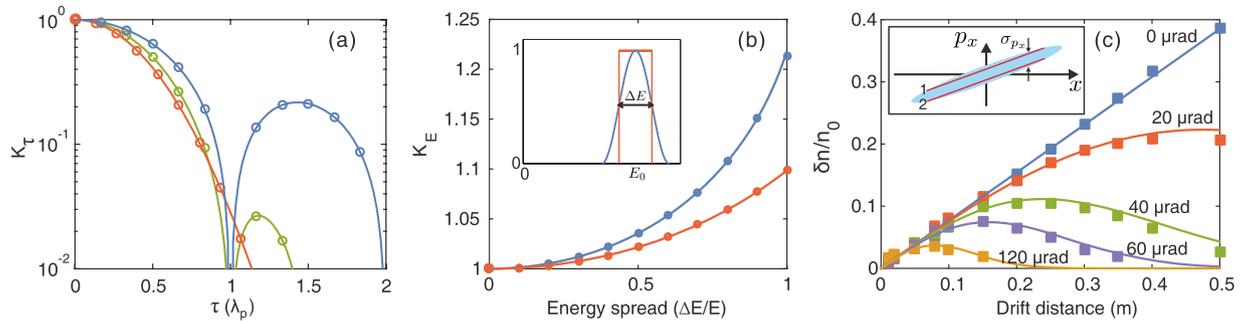

**Figure 5. Effects of probe length, energy spread and emittance.** (**a**) shows the correction factor $K_\tau$ for probes with different beam current profiles and lengths. (**b**) shows the correction factor $K_E$ for two probes with different energy spectra. (**c**) shows the density perturbation of probes with different emittance versus drift distance. In all figures, the solid lines represent theoretical predictions and the dots are for simulation results.

**Non-ideal probe beams.** The probes used in previous simulations (Figs 2–4) are ideal electron beams with zero energy spread and emittance, and very short pulse length (e.g., 6.8 fs). To evaluate the feasibility of this method in real experiments, the effects of pulse length, energy spread and emittance of a realistic electron probe must be considered. In this section we only present basic theoretical and simulation results and the derivation of the formulas are provided in the supplementary materials.

Since the wake has a phase velocity very close to $c$, the wakefield sampled by the front and the back part of the probe changes as the probe traverses through the wake. Thus the total density perturbation of the probe, which is a summation of the perturbations of different slices sampling different parts of the wakefield, will decrease as the probe becomes longer. For a probe with length of $\tau$ (defined as the full width at half maxima of the current profile, in unit of $\lambda_p$), the density perturbation will be $\delta n/n_0 = K_\tau(\delta n/n_0)_0$, where $(\delta n/n_0)_0$ is the density perturbation for the ideal probe. For linear wakes, $K_\tau$ is derived for probes with different current profiles and the results are shown in Fig. 5(a). The blue, green and red lines are theoretical predictions for probe with flat-top ($K_\tau = |\frac{\sin(\pi\tau)}{\pi\tau}|$), $\sin^2$ ($K_\tau = |\frac{\sin(2\pi\tau)}{2\pi\tau} - \frac{\sin(2\pi\tau)}{2\pi(2\tau-1)} - \frac{\sin(2\pi\tau)}{2\pi(2\tau+1)}|$) and Gaussian ($K_\tau = e^{-(k_p\sigma_\tau)^2/2}$, where $\sigma_\tau = \tau\lambda_p/2\sqrt{2\ln 2}$) current profile, respectively. Here $|\cdot|$ represents the absolute value. The circles are simulation results. For the flat-top probe, $K_\tau$ goes to zero when $\tau = 1$. This is because the fields seen by the front and the back part of the probe are exactly the opposite and cancel with each other. There is similar feature for the $\sin^2$ probe. In all three cases, the decrement of the density modulation caused by the finite probe length is less than 3.5% if the probe length is shorter than 0.1 plasma wavelength.

Energy spread also affects the probe density perturbation. For a probe with relative energy spread of $x = \Delta E/E_0$, the density perturbation will be $K_E(\delta n/n_0)_0$. Here $E_0$ is the central energy and $\Delta E$ is the FWHM energy spread, as shown in the inset of Fig. 5(b). Then we can derive that $K_E = x^{-1}\ln\frac{2+x}{2-x}$ for a probe with flat-top energy spectrum and $K_E = x^{-1}\int_{1-x}^{1+x}\cos^2\left(\frac{\pi(\alpha-1)}{2x}\right)\frac{1}{\alpha}d\alpha$ for a probe with $\sin^2$ energy spectrum. The red (blue) solid line in Fig. 5(b) shows $K_E$ for a probe with flat-top ($\sin^2$) energy spectrum. The dots are simulation results. It can be seen from Fig. 5(b) that the probe density perturbation increases as the energy spread becomes larger. However, this increment is still very small for even quite large energy spread. For both the flat-top and the $\sin^2$ cases, the increment of the probe density perturbation is smaller than 4% for a 50% energy spread.

The effect of the emittance of the probe is also considered. Comparing with the ideal probe case, the density perturbation for a probe with finite emittance will be $K_\varepsilon(\delta n/n_0)_0$, where $K_\varepsilon = e^{-(k_p\sigma_\theta L)^2/2}$, $k_p$ is plasma wave number, $L$ is the drift distance and $\sigma_\theta = \sigma_{p_x}/p_0$ is the relative slice momentum spread. The definition of σpx is given in the inset of Fig. 5(c), which is assumed to be constant for different x for simplicity. For beams generated from a LWFA with typical emittance of ~1 mmmrad, σθ will decrease to tens of μrad after the beam free drifts for tens of cm. The theoretical and simulation results for different σθ are shown in Fig. 5(c). The solid lines are for theoretical predictions (derived for linear wakes) and the dots are from simulation results. The blue line and blue dots show the density perturbation of an ideal probe, which is proportional to the drift distance $L$ according to Eq. (3). For probes with nonzero emittance, the density perturbation increases linearly when $L$ is small and decreases when $L$ becomes large. For a given drift distance $L$, the density perturbation decreases as σθ increases.

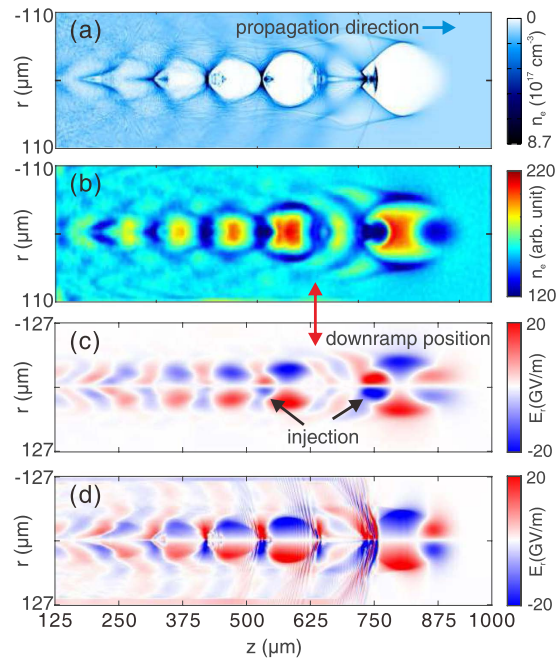

**Figure 6. Imaging of down ramp injection.** (**a**) Plasma density map in simulation. (**b**) Probe density on screen. The reconstructed and simulated $E_r$ fields are shown in (**c,d**), respectively. The red arrow shows the position where the density downramp starts. The black arrows in (**c**) indicate the field generated by the injected electron bunches. The drive bunch that generates the wake is not shown for clarity.

Another property of the probe beam that may affect the reconstruction accuracy is the local homogeneity of the probe. The validity of the theoretical model requires that the density perturbation induced by the wakefield $\delta n/n_0$ should be less than 1. As a result, a reasonable lateral intensity variation in the amplitude of the probe beam itself should better be less than 10%. It is worth noting that this variation is not for the entire transverse profile of the probe but for the local positions containing the wakefield information. Furthermore, the lateral intensity variations can be easily filtered out as long as its spatial frequency is not very close to the plasma frequency, for example, $>1.1\omega_p$ or $<0.9\omega_p$ if we assume that the uncertainty in determining the local plasma frequency from the measured density modulation is 10%. Such a probe beam can be readily obtained, especially when considering that the homogeneity of the probe transverse profile can be improved by scattering induced by passing the probe beam through a thin foil before probing the wakefield.

Finally, the density perturbation for a real probe that takes into account all the above effects can be rewritten as

$$\frac{\delta n}{n_0} = I_0 \cdot K_\tau \cdot K_E \cdot K_\varepsilon / M \quad (9)$$

where $I_0$ is the density perturbation of an ideal probe. The factors $K_\tau$, $K_E$ and $K_\varepsilon$ are corrections due to the finite probe length, energy spread and emittance, respectively. The last factor $M$ represents the geometrical magnification, which equals to unit for a parallel probe and $>1$ for diverging probes.

**Imaging of downramp injection.** The ultrashort relativistic electron probe opens the possibility of diagnosing many important phenomena in plasma based wakefield acceleration. In this section, we show an example of capturing snapshot of downramp injection[31] using FREP. The downramp injection is a promising controllable injection concept due to its potential on very localized high quality injection.

Figure 6(a) shows the plasma density map in the simulation. The plasma density decreases linearly from $1.1 \times 10^{17}\,\mathrm{cm^{-3}}$ to $7.8 \times 10^{16}\,\mathrm{cm^{-3}}$ in a distance of $100\,\mu\mathrm{m}$. As a result, an electron bunch is injected due to the expansion of the bubble. The details of the simulation are described in the Methods section. Figure 6(b) shows the snapshot of the wake, obtained by a 33 fs, 204 MeV electron probe. The density snapshot has basically an opposite phase when compared against the plasma density of the wake. The reconstructed and simulated transverse field $E_r$ are shown in Fig. 6(c,d), respectively. In both Fig. 6(b,c), the density transition regions can be seen clearly at the position indicated by the red arrow.

The wakefield is affected by the injected bunches due to their large charge densities and peak currents, which can be found readily in Fig. 6(c), as indicated by the black arrows. The $E_r$ field generated by the injected bunch in the first bubble is much more intense than the bunch in the second bubble, indicating that it has a higher charge density. This is to be expected because the first bunch is injected due to the density downramp by design, while the second bunch is self-injected due to the breaking of the highly nonlinear plasma wake. The simulated $E_r$ field is shown in Fig. 6(d) as a reference. It is once again clear that main features of the wake and the injection are qualitatively captured by the electron probe.

## Discussion

There are three time scales in FREP. The first is the wake evolving time, the second is probe transition time and the last is the pulse length of the probe. For typical wakes, especially for linear wakes, the evolving time is much longer than the transition time of the probe, which is also the assumption of our theoretical model. In this situation, the quality of field reconstruction depends on the probe pulse length. The snapshot of the wakefield structure will be clearer and the reconstruction will be more accurate if probe becomes shorter. Typically, the pulse length of electron bunches generated from LWFAs is a few fs[27,32], which is sufficiently short to revolve plasma wakes at a time scale of femtosecond.

To resolve the microscopic plasma wakes, spatial resolution as good as a few microns is required. In the method we proposed, the contributions of $E_z$ and $E_r$ field are decoupled by using the Panofsky-Wenzel theorem[29]. Therefore, the spatial resolution of the snapshot of wakefield structure is only limited by the screen that converting the electron density into light signals and the subsequent imaging system. By using thin (tens of $\mu$m) phosphor screens or OTR foils, spatial resolution as good as few microns is readily achievable[33].

In summary, we find that a few fs electron bunch containing few pC of charge with central energy on the order of 100 MeV ($\leq$20% energy spread and few mmmrad emittance) is suitable for diagnosing plasma based wakefield through theoretical analysis and supporting 3D PIC simulations. Such an electron bunch can be readily produced by a laser wakefield accelerator. Using this method, the electric field structure of a second plasma wake can be readily captured and reconstructed. Accurate longitudinal and transverse fields of linear wakes in low density plasmas can be obtained. Furthermore, important features (such as period, width and shape) and qualitative field information of nonlinear wakes can also be obtained, making this method very useful to diagnosing the nonlinear processes in plasma based wakefield acceleration. An example on capturing density downramp injection in a nonlinear wake is shown through 3D PIC simulations to illustrate the potential of this method.

## Methods

**Simulation.** The 3D PIC simulations are carried out using the code OSIRIS[28]. The dimensions for the simulation box are chosen as $1000 \times 400 \times 278\,\mu$m, divided into $1000 \times 400 \times 300$ cells along the $z$, $y$ and $x$ direction, respectively. In all simulations, a pre-ionized plasma with a uniform density profile is used. The peak density of the plasma is set to $1.1 \times 10^{17}\,\text{cm}^{-3}$. The wake is excited by an electron beam driver propagating along $z$ direction in all cases. The driver has a Gaussian profile in all three dimensions. The length and width of the driver are $20\,\mu$m ($\sigma_z$) and $17.8\,\mu$m ($\sigma_r$) respectively. For the linear case shown in Figs 2(a) and 3, the peak density of the driver is 5% of the plasma density. For the nonlinear case shown in Figs 2(c) and 4, the peak density of the driver equals to the plasma density. For the downramp injection case shown in Fig. 6, a density transition region is used. The plasma density decreases from $1.1 \times 10^{17}\,\text{cm}^{-3}$ to $7.8 \times 10^{16}\,\text{cm}^{-3}$ linearly in a distance of $100\,\mu$m. The peak density of the driver is increased to 2 times of the plasma density ($2.2 \times 10^{17}\,\text{cm}^{-3}$).

An electron probe is initialized close to one side of the simulation box. The distance between the probe center and the driver axis is $170\,\mu$m, which is large enough so that the wake amplitude is zero at the probe position. The probe is pushed along the $x$ direction after a proper time delay to sample the wakefield. The central energy of the probe is 204 MeV. The probe used for Figs 2–4 are ideal probes, i.e., without energy spread and emittance. The length of the probe is 6.8 fs, with a flat-top current profile. The probe has a rectangular shape ($995 \times 255\,\mu$m) and a uniform density profile along the perpendicular ($y$ and $z$) direction. The total charge of the probe is 14.4 pC. The probe used in the downramp case shown in Fig. 6 is realistic, with an energy spread of 10% and an emittance of $\varepsilon_{\text{geo}} = 2.5\,\mu$m. The duration of the probe is 33 fs, with a flat-top current profile. Other parameters are the same with probes used for Fig. 2.

In the simulations shown in Fig. 5(a), probes with different lengths traverse through a linear wake with amplitude of 5 GV/m, $\lambda_p = 60\,\mu$m and $k_p\sigma_E = 1$. The amplitude of the transverse momentum modulation, which is proportional to the density perturbation, is calculated along the wake propagation direction. Each circle in Fig. 5(a) represents the ratio of the amplitude of transverse momentum modulation between the realistic probe and the ideal probe.

In the simulations shown in Fig. 5(b), infinitely short probes with different energy spreads propagate transversely through a linear wake with amplitude of 5 GV/m, $\lambda_p = 76\,\mu$m and $k_p\sigma_E = 0.84$. Each dot in Fig. 5(b) represents the ratio of amplitude of the density perturbation between the realistic probe and the ideal probe.

In the simulations shown in Fig. 5(c), the plasma density is set to $3.1 \times 10^{17}\,\text{cm}^{-3}$ and the peak density of the driver is decreased to $3.5 \times 10^{15}\,\text{cm}^{-3}$. Probes with different transverse momentum (emittance) are initialized. Other parameters are the same as those in Fig. 2.

### Acknowledgements

This work was supported by the National Basic Research Program of China No. 2013CBA01501, NSFC Grant No. 11425521, No. 11535006, No. 11175102, No.11005063, No. 11375006 and No. 11475101, the Foundation of CAEP No. 2014A0102003, Tsinghua University Initiative Scientific Research Program, the Thousand Young Talents Program, DOE Grants No. DE-FG02-92-ER40727, No. DE-SC0008491, and No. DE-SC0008316, and NSF Grants No. PHY-145386, No. PHY-0936266, No. PHY-0960344, and No. ACI-1339893. Simulations are performed on Hoffman cluster at UCLA and Hopper cluster at National Energy Research Scientific Computing Center (NERSC).

### Author Contributions

C.J.Z. and W.L. proposed the concept. C.J.Z. developed the theoretical model and performed the simulations. C.J.Z., W.L., X.L.X. and W.B.M. discussed the theoretical model. C.J.Z., W.L. and C.J. wrote the paper. J.F.H., F.L., C.H.P., Y.W., Y.P.W. and Y.Q.G. contributed to refining the details of the paper. All authors reviewed the manuscript.

### Additional Information

**Supplementary information** accompanies this paper at http://www.nature.com/srep

**Competing financial interests:** The authors declare no competing financial interests.

**How to cite this article**: Zhang, C. J. *et al.* Capturing relativistic wakefield structures in plasmas using ultrashort high-energy electrons as a probe. *Sci. Rep.* **6**, 29485; doi: 10.1038/srep29485 (2016).